\documentclass[12pt,journal,a4paper,draftcls,oneside,onecolumn]{IEEEtran}

\usepackage[compress]{cite}
\usepackage{stmaryrd}
\usepackage{mathrsfs}
\usepackage{amsfonts}
\usepackage{amsmath}
\usepackage{cases}
\usepackage{txfonts}
\usepackage{amssymb}
\usepackage{epsfig}
\usepackage{pdfpages}
\usepackage{algorithm}
\usepackage{algpseudocode}
\usepackage{slashbox}
\usepackage{multirow}

\newcommand{\fat}[1]{\mbox{\boldmath$#1$}}

\begin{document}

	\title{Performance Analysis and Code Design for Resistive Random-Access Memory Using Channel Decomposition Approach}

	\author{{Guanghui Song, \IEEEmembership{Member, IEEE},
			Meiru Gao, 
			Ying Li, \IEEEmembership{Member, IEEE}, Bin Dai, \IEEEmembership{Member, IEEE},
			and
			Kui Cai, \IEEEmembership{Senior Member, IEEE}

		}
	
	\thanks{
		
		Guanghui Song, Meiru Gao, and Ying Li are with School of Communication Engineering, Xidian University, Xi'an, 710071, China (songguanghui@xidian.edu.cn, 22011210992@stu.xidian.edu.cn, yli@mail.xidian.edu.cn). 
		
		Bin Dai is with School of Internet of Things, Nanjing University of Posts and Telecommunications, Nanjing, 210094, China, and also with the Science,
		Mathematics and Technology Cluster, Singapore University of Technology
		and Design (SUTD), Singapore 487372 (daibin@njupt.edu.cn).
		
		Kui Cai is with Science, Mathematics and Technology Cluster, Singapore University of Technology and Design, Singapore, 487372 (cai\_kui@sutd.edu.sg).
			
	}
	}

	\maketitle

	\begin{abstract}
		A novel framework for performance analysis and code design is proposed to address the sneak path (SP) problem in resistive random-access memory (ReRAM) arrays. The main idea is to decompose the ReRAM channel, which is both
		non-ergodic and data-dependent, into multiple stationary memoryless channels. 
		A finite-length performance bound is derived by analyzing the capacity and dispersion of these stationary memoryless channels. 
		Furthermore, leveraging this channel decomposition, a practical sparse-graph code design is proposed using density evolution.  
		The obtained channel codes are not only asymptotic capacity approaching but also close to the derived finite-length performance bound. 
	\end{abstract}
	
	
	\section{Introduction}
	\IEEEPARstart{R}{e}sistive random-access memory (ReRAM) is a promising emerging non-volatile memory that utilizes memory cells arranged typically in a crossbar array to store information \cite{Strukov,Chen,Zahoor,Wong}. Each memory cell is essentially a memristor with two distinct resistance states, i.e., the High-Resistance State (HRS) with a resistance value of $R_0$ and the Low-Resistance State (LRS) with a resistance value of $R_1$. These states can be used to represent one bit of information, where HRS corresponds to a bit value of 0 and LRS to a bit value of 1. Consequently, an $N\times N$ memory array comprising $N^2$ memory cells is sufficient for storing an $N\times N$ binary data array $x^{N\times N}=[x_{m,n}]_{N\times N}$, where $x_{m,n}$ is either 0 or 1. 
	
		\begin{figure*}[t]
		\includegraphics[width=	5.8 in]{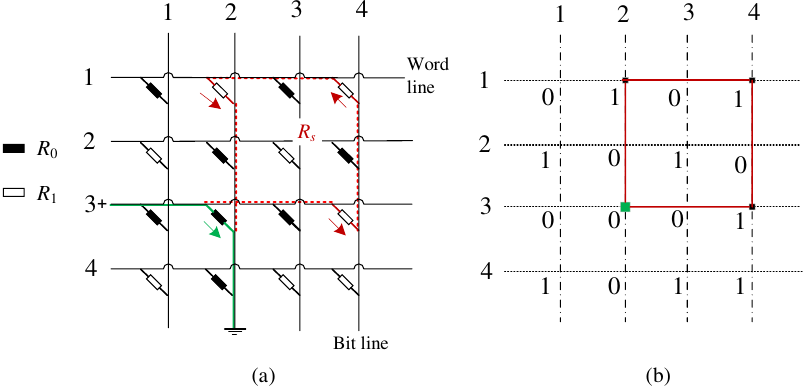}
		\centering
		\caption{(a) A $4\times4$ ReRAM array and (b) The corresponding data array. The green line plots the desired path when measuring the resistance of cell $(3, 2)$, and the red line forms an SP. Arrows show current flow directions. A reverse current flows across cell $(1, 4)$. } \label{fig:SPmodel}
	\end{figure*}
	
\begin{figure}[t]
		\includegraphics[width=
		3.8 in]{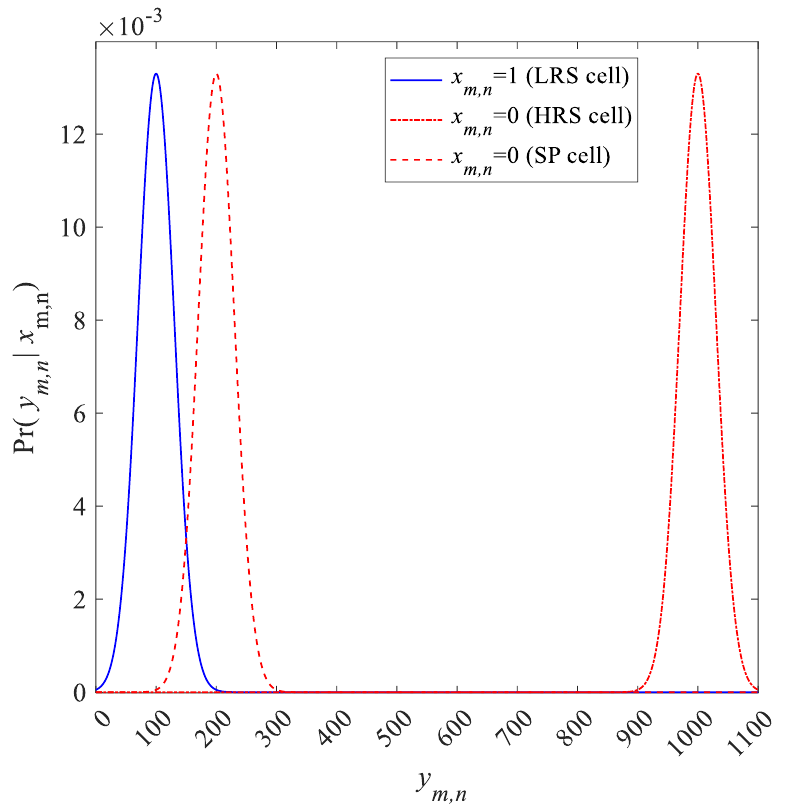}
		\centering
		\caption{PDF of the detected resistance value $y_{m,n}$ when the cell is an LRS cell ($x_{m,n}=1$), a HRS cell ($x_{m,n}=0$ without SP interference), or an SP cell ($x_{m,n}=0$ with SP interference). The associated resistance values are $R_0=1000\ \Omega, R_1=100\ \Omega$,  and $R_s=250\ \Omega$, and the noise  standard deviation is $\sigma=30$.} \label{fig:yPDF}
	\end{figure}

	Sneak path (SP) is an undesired current path during memory reading which severely degrades the data detection performance, \cite{Zidan,Naous,Luo1,Kvatinsky,Yaakobi}, as illustrated in Fig.~\ref{fig:SPmodel}. When a memory cell is read, a voltage is applied across the corresponding word line and bit line, causing current to flow through the cell. An SP represents an alternative current path that runs parallel to the intended path for resistance measurement. This alternative path forms a closed loop, originating and terminating at the target cell being read, while traversing through LRS cells via alternating vertical and horizontal steps.
	Fig.~\ref{fig:SPmodel} depicts an SP during the reading of cell $(3, 2)$. The green line indicates the desired path for resistance measurement, while the red line, passing through cells $(3, 4)\rightarrow(1, 4)\rightarrow(1, 2)$, represents an SP. In such scenarios, the detected resistance value (in the absence of noise) becomes
	\begin{equation}
	R_0^\prime=\left(\frac{1}{R_0}+\frac{1}{R_s}\right)^{-1}
	\end{equation}
	where $R_s$ represents the parasitic resistance introduced by the SP.  Typically, we have $R_0>>R_0^\prime>R_1$. Since the SP always decreases the measured resistance value, it is harmful only when a cell with HRS is read. 
	Fig.~\ref{fig:yPDF} depicts the probability density function (PDF) of the readback signal $y_{m,n}$, which represents the detected resistance value contaminated with Gaussian noise, for memory cell $(m,n)$. This illustration considers scenarios where the stored data is either $x_{m,n}=1$ or $x_{m,n}=0$, with and without the interference of an SP. We adhere to the parameter setting employed in previous works \cite{Ben,CZH,SongTcom2021,SongTcom2022}, where the respective resistance values are designated as $R_0=1000\ \Omega, R_1=100\ \Omega$,  and $R_s=250\ \Omega$.  Notably, a significant cross-error probability is observed in the presence of an SP. In the following, cell $(m,n)$ is referred to as an LRS cell if 
	$x_{m,n}=1$; as an HRS cell if $x_{m,n}=0$ and the cell is not affected by an SP; and as an SP cell is $x_{m,n}=0$ and the cell is affected by an SP.

	The most commonly adopted method to mitigate SPs is the utilization of cell selectors, which are electrical devices that permit current flow in only one direction across memory cells. One example of this is the one-diode one-resistor (1D1R) type ReRAM, which incorporates a diode in series with each memory cell. Since an SP inherently generates a reverse current in at least one of the cells (e.g. cell $(1, 4)$ in Fig.~\ref{fig:SPmodel}) along the SP, if all selectors are functioning correctly, SPs can be completely prevented. However, selectors are also prone to errors due to
	imperfections in the production or maintenance of the memory leading to the reoccurrences of SPs, as reported in \cite{Ben,CZH,SongTcom2021,SongTcom2022}. 
	Therefore, $(m,n)$ is referred to as an SP cell if and only if the following three conditions (\textit{SP conditions}) are met: 
	
	1) $x_{m, n}=0$.
	
	2) There exists $i, j\in\{1,..., N\}$ that
	$x_{m, j}=x_{i, j}=x_{i, n}=1$.
	
	3) The selector located at cell $(i, j)$ is faulty (a reverse current flowing through this cell).\\
	Based on previous works \cite{CZH,SongTcom2021,SongTcom2022}, the occurrence of selector failure (SF) is random and is unknown to the data detector. According to conditions 2) and 3), an SF at position $(i,j)$ can only result in SPs if the cell at this position is an LRS cell, meaning $x_{i, j}=1$. Otherwise, the failure will have no impact on the array. Therefore, an SF occurring at an LRS cell is referred to as an \emph{active SF}.   
	Furthermore, the SP conditions above specifically constrain SPs of length 3, traversing exactly three HRS cells. Longer SPs are ignored due to their insignificant impact compared to those of length 3.
	Consistent with previous works \cite{CZH,SongTcom2021,SongTcom2022}, we do not take into account the superposition effect of multiple SPs. A more sophisticated SP model was explored in \cite{Ben}.

	Recent works \cite{Ben,CZH,SongTcom2022,SongTcom2021,Panpan1,Panpan2,SunTcom,DaiBin,Yuval,Thanh,SongTIT2023} addressed the SP problem using information theory and channel coding methods. 
	For uncoded ReRAM, works \cite{Ben,CZH,SongTcom2022} proposed efficient data detection schemes.  Ben-Hur and Cassuto \cite{Ben} presented a cell-independent data detection scheme without considering the inter-cell correlation. Chen \emph{et al.} \cite{CZH} and our previous work \cite{SongTcom2022} proposed joint data and SP detection schemes. In \cite{SongTcom2022}, we first detected the locations of SFs, based on which, the SPs and data were jointly detected. With an assumption of less than 3 SFs in the array, our scheme yields a near-optimal performance. 
	
	To further enhance the data storage reliability, channel codes including low-density parity-check (LDPC) and polar codes were applied to ReRAM \cite{SongTcom2021,Panpan1,Panpan2,SunTcom,DaiBin}. Regarding the code design,  a significant challenge arises from the non-ergodic nature of the ReRAM channel as shown in \cite{SongTIT2023}. In this scenario, due to the data dependency of SPs, the channel's performance fluctuates for each data array, especially when the array size is small. The channel was characterized as time-varying in \cite{SongTcom2021}. However, most previous works have overlooked this factor during code construction, instead designing the code based on an averaged channel performance.

	In this work, we propose a novel framework for performance analysis and code design for ReRAM. The main idea is to decompose the ReRAM channel, which is 
	non-ergodic and data-dependent, into multiple stationary memoryless (data-independent) channels. 
	By analyzing the capacity and dispersion of these stationary memoryless channels, we establish a finite-length performance bound, serving as a benchmark for the code design. To calculate this bound, we further analyze the probability distribution of the SP occurring rate within a memory array of a given size. This channel decomposition approach offers a clear understanding of non-ergodic channels
	and a new path for the code design.
	Moreover, leveraging this channel decomposition, we propose a practical sparse-graph code design utilizing density evolution (DE).  We demonstrate the Gaussanity of the log-likelihood ratio (LLR)  messages, based on which we develop a low-complexity two-dimensional DE for code optimization.   
	The obtained channel codes are not only asymptotic capacity approaching but also close to the derived finite-length performance bound. 
	
	The rest of this paper is organized as follows. Sec.~\ref{sec:model} introduces the coded ReRAM system model. In Sec.~\ref{sec:main}, we propose the channel decomposition and conduct the finite-length performance analysis.  Sec.~\ref{sec:threshold} presents the sparse-graph code design and develops the two-dimensional DE framework. Numerical results of code optimization and simulations are presented in Sec.~\ref{sec:numerical}. Finally, we conclude our work in Section~\ref{sec:end}.
	
	\section{Coded ReRAM System Model}\label{sec:model}

\begin{figure*}[t]
		\includegraphics[width=
		5.6 in]{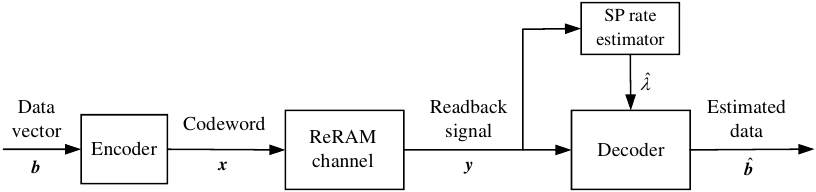}
		\centering
		\caption{System model of Coded ReRAM.} \label{fig:Coding_system}
	\end{figure*}

	The system model of coded ReRAM is illustrated in Fig.~\ref{fig:Coding_system}.
	In this model, data vector $\fat{b}=(b_1, b_2,...,b_{N^2R})$ is encoded into an $N\times N$ codeword array $\fat{x}=[x_{m,n}]_{N\times N}$ where $x_{m,n}\in\{0, 1\}$ and $R$ represents the code rate.  This encoded array $\fat{x}$ can then be stored in an $N\times N$ ReRAM array. Let 
	\begin{equation}
	R_x(e)\triangleq\left(\frac{1}{R_x}+(1-x)\frac{e}{R_s}\right)^{-1},\ \ \textrm{for} \ x, e\in\{0, 1\}.\label{eq:Rx}
	\end{equation}
	Thus, $R_1(0)=R_1(1)=R_1$ and $R_0(0)=R_0, R_0(1)=R_0^\prime$. Let $\Psi$ be the set that includes all the indices of cells with SFs in the array.  According to the SP conditions, the readback signal $\fat{y}=[y_{m,n}]_{N\times N}$ is written as
	\begin{equation}
	y_{m,n}=R_{x_{m,n}}\left(\bigcup_{(i,j)\in\Psi}x_{m,j}x_{i,j}x_{i,n}\right)+z_{m,n},\label{eq:Y}
	\end{equation}
	where $\bigcup$ is the logical OR operator, i.e., $\bigcup_{(i,j)\in\Psi}x_{m,j}x_{i,j}x_{i,n}=1$ if at least one of $(i,j)\in\Psi$ with $x_{m,j}x_{i,j}x_{i,n}=1$, otherwise, $\bigcup_{(i,j)\in\varphi}x_{m,j}x_{i,j}x_{i,n}=0$, and $z_{m,n}\sim \mathcal{N}(0, \sigma^2), m=1,...,N, n=1,...,N$ are independent and identically distributed (i.i.d.) samples of Gaussian noise.
	
	An LLR for each coded bit is calculated to facilitate the soft decoding. Different from the conventional binary input Gaussian channel, where the data bits 0 and 1 are modulated to two distinct voltage values, in \eqref{eq:Y}, if $x_{m,n}=0$, the resistance value of cell $(m,n)$ may vary as $R_0$ or $R_0^\prime$ depending on whether the cell is affected by an SP. Assuming Bernoulli $(q)$ data storage where Pr$(x_{m,n}=1)=q$ and Pr$(x_{m,n}=0)=1-q$ with$0<q<1$, the LLR for 
	$x_{m,n}$ is 
	\begin{align}
	L(x_{m,n}|& y_{m,n}) =\log\frac{(1-q)\textrm{Pr}(y_{m,n}|x_{m,n}=0)}{q\textrm{Pr}(y_{m,n}|x_{m,n}=1)}\nonumber\\
	=\log &\frac{\lambda \phi(y_{m,n}-R_0^\prime)+(1-\lambda)\phi(y_{m,n} - R_0)}{\phi(y_{m,n}-R_1)}+\log\frac{1-q}{q}\label{eq:LLR}
	\end{align}
	where $\lambda, 0\leq \lambda\leq 1$, is the probability that cell $(m,n)$ is an SP cell, given that $x_{m,n}=0$, and $\phi(z)\triangleq{1}/{(\sqrt{2\pi}\sigma)}e^{-\frac{z^2}{2\sigma^2}}$ is a Gaussian PDF with mean 0 and variance $\sigma^2$. 
	
	Strictly speaking, $\lambda$ is cell-specific and its value is related to the data pattern of the whole array $\fat{x}$. Thus, the optimal decoder should process the memory cells jointly, but the complexity will be very high \cite{SongTcom2022}. A simpler decoder that facilitates practical realization is carried out by treating the memory cell independently \cite{SongTcom2021,Panpan1,Panpan2,SunTcom,DaiBin}.  In this scenario, $\lambda$ is assumed to be array-specific but within the same array, a common $\lambda$ is adopted for all memory cells. The parameter $\lambda$ is referred to as the \textit{SP rate} of the array, reflecting the fraction of SP cells among the cells that are intended to store the data 0. Obviously, the SP rate depends on both the number of active SFs and the data pattern of the array. In practice, $\lambda$ is first inferred based on the readback signal array before decoding. Following previous works \cite{SongTcom2021,Panpan1,Panpan2,SunTcom,DaiBin}, $\lambda$ is estimated based on the hard decision of the readback signals. Specifically, let
	\begin{equation}
	\bar{y}_{m,n}=\arg\min_{r\in\{R_0^\prime,R_0,R_1\}}|y_{m,n}-r|\nonumber.
	\end{equation}
	The estimation of $\lambda$ is
	given by 
	\begin{align}\label{eq:channel_est}
	\hat{\lambda}=\frac{n_{R_0^\prime}}{n_{R_0^\prime}+n_{R_0}},\ n_r\triangleq\sum_{m=1}^N\sum_{n=1}^N1\{\bar{y}_{m,n}=r\}
	\end{align}
	where $1\{\mathcal{E}\}$ is the indication function which takes the value $1\{\mathcal{E}\}=1$ if $\mathcal{E}$ is satisfied; otherwise, it takes value $1\{\mathcal{E}\}=0$.

	\section{Channel Decomposition and Finite-Length Analysis}\label{sec:main}
	
	The SP rate affects the channel quality significantly. Furthermore, it depends on the number of active SFs in the array and the data pattern, both of which are random. This poses a significant challenge to code design, as channel codes are typically optimized for stationary memoryless channels.  Designing a code based on the average channel SP rate risks high decoding failure probability when the actual SP rate surpasses this average. On the contrary, designing a code based on the worst-case scenario results in a notably low code rate. In this section, we propose a channel decomposition approach that decomposes the ReRAM channel into multiple stationary memoryless channels, each with a distinct SP rate. By designing a code tailored to one of these stationary memoryless channels, we can achieve a specific rate with an outage probability. We analyze the probability distribution of the SP rate and the second-order rate of the sub-channels for a finite length of $N$ and derive a finite-length performance bound.  This bound serves as a benchmark for code design.

\begin{figure}[t]
		\includegraphics[width=
		3.6 in]{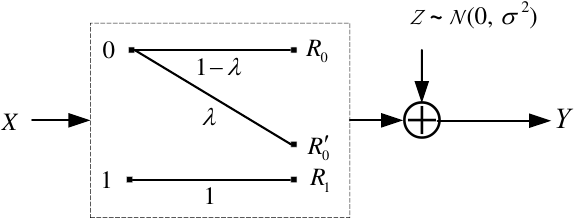}
		\centering
		\caption{$\lambda$-Gaussian channel model.} \label{fig:ESchannel}
	\end{figure}

	We define a $\lambda$-Gaussian channel as a stationary and memoryless channel characterized by binary input and continuous output, as illustrated in Fig.~\ref{fig:ESchannel}. This $\lambda$-Gaussian channel was also  termed the $(\lambda,\sigma)$-channel in \cite{SongTcom2021}. When  $\lambda=0$ or $\lambda=1$, the $\lambda$-Gaussian channel reduces to a binary-input additive white Gaussian noise (BIAWGN) channel. 
	
	For a given array size of $N\times N$, if the data input distribution and the SF probability distribution jointly yield a probability distribution of the SP rate, denoted as $F_N(\lambda), 0\leq\lambda\leq 1$, we say that the ReRAM channel can be decomposed into 
	a set of $\lambda$-Gaussian channels with a profile of $F_N(\lambda)$. In other words, the ReRAM channel is a mixture of multiple $\lambda$-Gaussian channels, where $F_N(\lambda)$ specifies the proportion of each constituent channel within the mixture.  
	Let $C_{\lambda}$ and 
	$V_{\lambda}$ be the channel capacity and dispersion of the $\lambda$-Gaussian channel. According to the Polyanskiy-Poor-Verdu (PPV) bound \cite{polyanskiy}, the minimum word error probability achievable using a code with rate $R$ and length $N^2$ over the ReRAM channel can be approximated by 
	\begin{align}
	P_\text{error}\approx \sum_{\lambda}F_N(\lambda)Q\left(\frac{N(C_{\lambda}-R)}{\sqrt{V_{\lambda}}}\right) \label{eq:bound}
	\end{align}
	where $Q(x)=\int_{x}^\infty{1}/{\sqrt{2\pi}}e^{-\frac{u^2}{2}}du$.
	
	In the rest of this section, we provide approximation analyses for $F_N(\lambda), C_{\lambda}$, and $V_{\lambda}$. In our analyses, we assume Bernoulli $(q)$ data storage and with a fixed number of $K (K<<N)$ active SFs in the array. Under this assumption, the distribution of $\lambda$ depends solely on the randomness of the data pattern. However, this does not affect the generality of our analyses. For the scenarios involving a random number of SFs, we can first average $F_N(\lambda)$ over the probability distribution of the SF number and then apply the bound given in \eqref{eq:bound}.   
	
	%
	
	\begin{figure*}[t]
		\includegraphics[width=
		5.4 in]{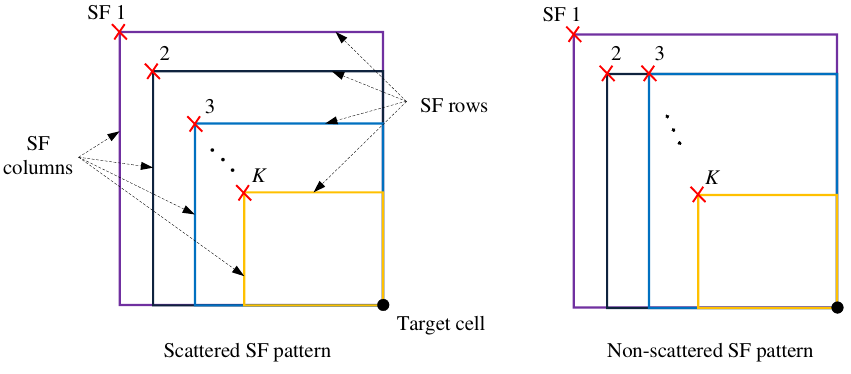}
		\centering
		\caption{An example of scattered and non-scattered SF patterns.} \label{fig:Scattered}
	\end{figure*} 	
	
	\subsection{Probability Distribution of SP Rate $F_N(\lambda)$}
	Considering Bernoulli $(q)$ data storage with $K$ active SFs in an $N\times N$ ReRAM array, we analyze the probability distribution of the SP rate $\lambda_N$. We assume a type of scattered SF pattern where the SFs are located in different rows and columns as shown in Fig.~\ref{fig:Scattered}. The analysis can be extended to the case of a non-scattered SF pattern. We first show the convergence of SP rate as $N\rightarrow\infty$ by analyzing its mean and variance. Then we propose a Gaussian approximation for $F_N(\lambda)$.  
	
	We show that if the data is i.i.d. Bernoulli $(q)$, as $N\rightarrow\infty$,  with probability 1,
	\begin{equation}
	\lambda_N\rightarrow  1-\left(1-\frac{2K}{N}\right)(1-q^2)^K-\frac{2K}{N}(1-q^2)^{K-1}+o\Big(\frac{1}{N}\Big)\label{eq:Elam}
	\end{equation}
	where we neglected higher-order terms of $o(1/N)$ but kept the $O(1/N)$ terms to ensure that our analysis is relevant to the finite block-length situation.  
	In the following, we demonstrate that the expectation of the SP rate, $E[\lambda_N]$, tends to the RHS of \eqref{eq:Elam}, and its variance $D[\lambda_N]\rightarrow O(1/N)$ as $N\rightarrow\infty$. Thus, given $\epsilon>0$, Chebyshev's inequality yields 
	\begin{align}
	\text{Pr}\left(\big|\lambda_N-E[\lambda_N]\big|\geq\epsilon\right)\leq \frac{D[\lambda_N]}{\epsilon^2}\rightarrow 0.\nonumber
	\end{align}
	
	\subsubsection{ $E[\lambda_N]$}
	Let $X_\text{sf}$ be the set containing the indices of memory cells that belong to the SF rows and columns and $X_\text{nsf}$ be the set containing the indices of the remaining memory cells. Here an SF row or column refers to the row or column that contains an active SF, as illustrated in Fig.~\ref{fig:Scattered}. We divide the data array $\fat{x}$ into two parts: $\{x_{m,n}, (m,n)\in X_\text{sf}\}$ and $\{x_{m,n}, (m,n)\in X_\text{nsf}\}$.  For ReRAM with $K$ scattered active SFs, the cardinality of these two sets are
	\begin{align}
	|X_\text{sf}|&=2KN-K^2\nonumber\\
	|X_\text{nsf}|&=N^2-2KN+K^2\nonumber.
	\end{align}
	
	Since the data in $\fat{x}$ is i.i.d., according to Chebyshev's inequality, 
	\begin{equation}
	\text{Pr}\left(\Big |\sum_{m=1}^N\sum_{n=1}^N1\big\{x_{m,n}=0\big\}-(1-q)N^2\Big |\geq\epsilon\right)\leq \frac{q-q^2}{\epsilon^2N^2}\nonumber
	\end{equation}
	the number of 0's in the array converges to $(1-q)N^2$ rapidly as $N$ increases.  From \eqref{eq:Y}, cell $(m,n)$ will be affected by SP if and only if  $x_{m,n}=0$ and $\bigcup_{(i,j)\in\Psi^*}x_{m,j}x_{i,n}=1$, where $\Psi^*$ is the set containing the indices of all the active SFs. 
	Thus, the SP rate of the array is derived as
	\begin{eqnarray}
	\lambda_N\!\!\!\!\!\!\!\!\!&&\rightarrow\frac{1}{(1-q)N^2}\sum_{m=1}^N\sum_{n=1}^N1\Bigg\{\bigcup_{(i,j)\in\Psi^*}x_{m,j}x_{i,n}=1, x_{m,n}=0\Bigg\}\nonumber\\
	\!\!\!\!\!\!\!\!\!&&=\frac{1}{(1-q)N^2}\sum_{(m,n)\in X_\text{nsf}}1\Bigg\{\bigcup_{(i,j)\in\Psi^*}x_{m,j}x_{i,n}=1, x_{m,n}=0\Bigg\}\nonumber\\
	\!\!\!\!\!\!\!\!\!&&\ \ \ \ \ \ \ \ \ + \frac{1}{(1-q)N^2}\sum_{(m,n)\in X_\text{sf} \atop (m,n)\notin\Psi^*}1\Bigg\{\bigcup_{(i,j)\in\Psi^* \atop i\neq m, j\neq n}x_{m,j}x_{i,n}=1, x_{m,n}=0\Bigg\}\nonumber
	\end{eqnarray}
	where the second term of the above expression can be explained as follows: first $(m,n)\in\Psi^*$ is in conflict with the condition $x_{m,n}=0$; second, if either $i=m$ or $j=n$, the product $x_{m,j}x_{i,n}=1$ would contradict the condition $x_{m,n}=0$.
	Using $1\{\bigcup_{j}x_j=1, x=0\}=(1-\prod_{j}(1-x_j))(1-x)$, after neglecting the high order terms of $o(1/N)$, the above equation is rewritten as 
	\begin{align}
	\lambda_N&\rightarrow\left(1-\frac{2K}{N}\right)A_N+\frac{2K}{N}B_N+o\Big(\frac{1}{N}\Big)\label{eq:lamAB}
	\end{align}
	where
	\begin{align}
	A_N=&\frac{1}{|X_\text{nsf}|(1-q)}\sum_{(m,n)\in X_\text{nsf}}(1-x_{m,n})\Bigg (1-\prod_{(i,j)\in\Psi^*}(1-x_{m,j}x_{i,n})\Bigg)\nonumber\\
	B_N=&\frac{1}{(|X_\text{sf}|-K)(1-q)}\sum_{(m,n)\in X_\text{sf}\atop (m,n)\notin\Psi^*}(1-x_{m,n})\Bigg (1-\prod_{(i,j)\in\Psi^* \atop i\neq m, j\neq n}(1-x_{m,j}x_{i,n})\Bigg).\nonumber
	\end{align}
	
	Using the i.i.d. assumption of data, we have 
	\begin{align}
	E[A_N]&=\frac{1}{|X_\text{nsf}|(1-q)}\sum_{(m,n)\in X_\text{nsf}}(1-E[x_{m,n}])\nonumber\\
	&\ \ \ \ \ \ \ \ \    \times \Bigg (1-\prod_{(i,j)\in\Psi^*}(1-E[x_{m,j}]E[x_{i,n}])\Bigg)\nonumber\\
	&=1-(1-q^2)^K\label{eq:meanlam}
	\end{align}
	and $E[B_N]\rightarrow1-(1-q^2)^{K-1}$, where we used $|\Psi^*|=K$. Therefore,
	\begin{align}
	E[\lambda_N]&\rightarrow1-\left(1-\frac{2K}{N}\right)(1-q^2)^K-\frac{2K}{N}(1-q^2)^{K-1}+o\Big(\frac{1}{N}\Big).\nonumber
	\end{align}
	
	Note that the scattered SF pattern represents the worst-case scenario. For the non-scattered SF pattern shown in Fig.~\ref{fig:Scattered}, the expected SP rate, when neglecting the $o(1)$ terms, will be $1-(1-2q^2+q^3)(1-q^2)^{K-2}$. This value is smaller than the right-hand side (RHS) of \eqref{eq:meanlam}.

	\subsubsection{ $D[\lambda]$}
	Based on \eqref{eq:lamAB}, we have 
	\begin{align}
	\!\!\!\!\! D[\lambda_N]\rightarrow &\left(1-\frac{2K}{N}\right)^2D[A_N]+ \left(\frac{2K}{N}\right)^2D[B_N]\nonumber\\
	& +\frac{4K}{N}\left(1-\frac{2K}{N}\right)E\Big[\big(A_N-E[A_N]\big)\big(B_N-E[B_N]\big)\Big].\label{eq:Dlamterm}
	\end{align}
	We analyze $D[A_N]$ and then show that the remaining terms are all $o(1/N)$.
	
	Once again, utilizing the i.i.d. assumption of data, we have 
	\begin{align}
	E\left [ A_N^2 \right ] & =  \frac{1}{|X_\text{nsf}|^2(1-q)^2}\!\sum_{(m,n)\in X_\text{nsf}}\sum_{(m^\prime,n^\prime)\in X_\text{nsf}}\!\!E\Big[(1-x_{m,n})(1-x_{m^\prime,n^\prime})\Big]\nonumber\\
	&  \times E\Bigg[\Big(1-\prod_{(i,j)\in\Psi^*}(1-x_{m,j}x_{i,n})\Big) \Big(1-\prod_{(i,j)\in\Psi^*}(1-x_{m^\prime,j}x_{i,n^\prime})\Big)\Bigg]\nonumber.
	\end{align}
	We simplify the above expression as follows.  
	First, the value inside the first summation of $\sum_{(m,n)\in X_\text{nsf}}$ is independent 
	of the index $(m,n)$. Second, the second expectation can be further expanded as 
	\begin{align}
	&E\Bigg[\Big(1-\prod_{(i,j)\in\Psi^*}(1-x_{m,j}x_{i,n})\Big) \Big(1-\prod_{(i,j)\in\Psi^*}(1-x_{m^\prime,j}x_{i,n^\prime})\Big)\Bigg]\nonumber\\
	=& 1-2(1-q^2)^K+\Big(E_{m^\prime,n^\prime}\Big)^K\nonumber
	\end{align} 
	where
	\begin{align}
	E_{m^\prime,n^\prime}= E\Big[(1-x_{m,j}x_{i,n})(1-x_{m^\prime,j}x_{i,n^\prime})\Big]\nonumber.
	\end{align}
	Thus, we have the following simplified expression
	\begin{align}
	E\left [ A_N^2 \right ]  =&\  \frac{1}{|X_\text{nsf}|(1-q)^2}\sum_{(m^\prime,n^\prime)\in X_\text{nsf}}E\Big[(1-x_{m,n})(1-x_{m^\prime,n^\prime})\Big]\nonumber\\
	&\ \ \ \ \ \ \ \times \Big(1-2(1-q^2)^K+\Big(E_{m^\prime,n^\prime}\Big)^K\Big)\nonumber.
	\end{align}

	For $(m,n)\in X_\text{nsf}$, we partition the set $X_\text{nsf}$ into four subsets:
	\begin{align}
	S_1&=\Big\{(m^\prime,n^\prime)|(m^\prime,n^\prime)=(m,n)\Big\}\nonumber\\
	S_2&=\Big\{(m^\prime,n^\prime)\Big| m^\prime\neq m, n^\prime\neq n, (m^\prime,n^\prime)\in X_\text{nsf}\Big\}\nonumber\\
	S_3&=\Big\{(m,n^\prime)\big| n^\prime\neq n, (m,n^\prime)\in X_\text{nsf}\Big\}\nonumber\\
	S_4&=\Big\{(m^\prime,n)\big| m^\prime\neq m, (m^\prime,n)\in X_\text{nsf}\Big\}\nonumber. 
	\end{align}
	It is obvious that $S_i, i=1,2,3,4$, is a division of $X_\text{nsf}$. 
	Their cardinalities are 
	\begin{align}
	|S_1|&=1\nonumber\\
	|S_2|&=N^2-2(K+1)N+(K+1)^2\nonumber\\
	|S_3|&=|S_4|=N-(K+1)\nonumber.
	\end{align}
	Firstly, discarding the term $(m^\prime,n^\prime)\in S_1$ does not impact the limitation as it is of $O(1/N^2)$. Moreover, for $(m^\prime,n^\prime)\notin S_1$,  
	\begin{equation}
	E\big[(1-x_{m,n})(1-x_{m^\prime,n^\prime})\big]=(1-q)^2.\nonumber\\
	\end{equation}
	Leveraging the symmetry of entries in $S_3$ and $S_4$, we obtain 
	\begin{align}
	E\left [ A_N^2 \right ]  &\rightarrow \! 1-2(1-q^2)^K+ \frac{1}{|X_\text{nsf}|}\Bigg[\sum_{(m^\prime,n^\prime)\in S_2}\!+2\sum_{(m^\prime,n^\prime)\in S_3}\Bigg] (E_{m^\prime,n^\prime})^K.\nonumber
	\end{align}

	Since for $(m^\prime,n^\prime)\in S_2$, $(1-x_{m,j}x_{i,n})$ and $(1-x_{m^\prime,j}x_{i,n^\prime})$ are independent, we have 
	\begin{align}
	E_{m^\prime,n^\prime}&= E\Big[1-x_{m,j}x_{i,n}\Big]E\Big[1-x_{m^\prime,j}x_{i,n^\prime}\Big]\nonumber\\
	&=(1-q^2)^2\nonumber.
	\end{align}
	For $(m^\prime,n^\prime)\in S_3$, we have  
	\begin{align}
	E_{m^\prime,n^\prime}&= E\Big[(1-x_{m,j}x_{i,n})(1-x_{m,j}x_{i,n^\prime})\Big]\nonumber\\
	&= E\Big[1-x_{m,j}x_{i,n}-x_{m,j}x_{i,n^\prime}+x_{m,j}x_{i,n}x_{i,n^\prime}\Big]\nonumber\\
	&= 1-2q^2+q^3.\nonumber
	\end{align}
	Utilizing the facts: 
	\begin{align}
	\frac{|S_2|}{|X_\text{nsf}|}=&\frac{N^2-2(K+1)N+(K+1)^2}{N^2-2KN+K^2}\nonumber\\
	\rightarrow & 1-\frac{2}{N}+o\Big(\frac{1}{N}\Big) \nonumber
	\end{align}
	\begin{align}
	\frac{|S_3|}{|X_\text{nsf}|}=\frac{N-(K+1)}{N^2-2KN+K^2}
	\rightarrow  \frac{1}{N}+o\Big(\frac{1}{N}\Big) \nonumber
	\end{align}
	And through some mathematical manipulations, we obtain
	\begin{align}
	E\left [ A_N^2 \right ] \rightarrow \Big(1-(1-q^2)^K\Big)^2+\frac{2}{N}\Big(\big(1-2q^2+q^3\big)^K-(1-q^2)^{2K}\Big)\nonumber.
	\end{align}
	Here we have omitted higher-order terms of $o(1/N)$. 
	Therefore, 
	\begin{align}
	D[ A_N]&=E\left [A_N^2 \right ]-E\left [A_N \right ]^2\nonumber\\
	&\rightarrow \frac{2}{N}\Big(\big(1-2q^2+q^3\big)^K-(1-q^2)^{2K}\Big)+o\Big(\frac{1}{N}\Big)\nonumber.
	\end{align}
	
						\begin{figure*}[t]
		\includegraphics[width=
		6.3 in]{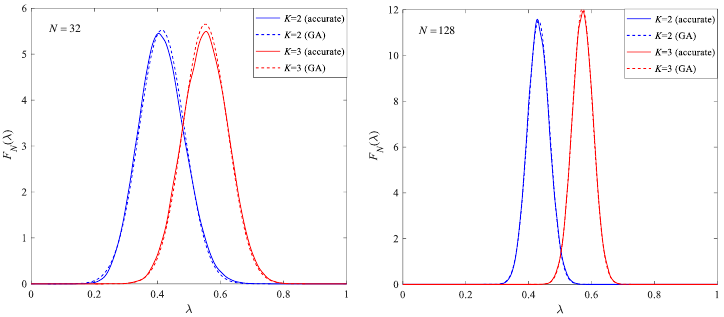}
		\centering
		\caption{Accurate probability distribution $F_N(\lambda)$ of SP rate and their Gaussian approximation (GA) when $q=1/2$.} \label{fig:LamGA}
	\end{figure*}

	It leaves us to show that the remaining terms of \eqref{eq:Dlamterm}  are all $o(1/N)$.
	Since $D[A_N]\rightarrow0$ as $N\rightarrow\infty$, we have that $A_N\rightarrow E[A_N]$ in probability.
	Following the same procedure, we can show that $D[B_N]\rightarrow0$ as $N\rightarrow\infty$ and thus, $B_N\rightarrow E[B_N]$ in probability. Since both $A_N$ and $B_N$ are bounded, i.e., $A_N\leq1/(1-q), B_N\leq1/(1-q)$, we have $E\Big[\big(A_N-E[A_N]\big)\big(B_N-E[B_N]\big)\Big]\rightarrow0$. Therefore, all the remaining terms of \eqref{eq:Dlamterm} are $o(1/N)$. Finally, we obtain
	\begin{align}
	D[\lambda_N]\rightarrow \frac{2}{N}\Big(\big(1-2q^2+q^3\big)^K-(1-q^2)^{2K}\Big)+o\Big(\frac{1}{N}\Big)\label{eq:Dlam}.
	\end{align}

	\subsubsection{Gaussian Approximation of $F_N(\lambda)$}
	We have observed that $F_N(\lambda)$ is very close to the Gaussian distribution of $\mathcal{N}(E[\lambda_N], D[\lambda_N])$, where the higher-order terms of $o(1/N)$ in \eqref{eq:Elam} and \eqref{eq:Dlam} are neglected. Fig.~\ref{fig:LamGA} shows that Gaussian approximation is quite accurate even when the array size $N$ is as small as  $32$.
	Thus, we use Gaussian PDF as an approximation of $F_N(\lambda)$.

	\subsection{Channel Capacity and Dispersion of $\lambda$-Gaussian Channel}

	In this section, we analyze the channel capacity $C_\lambda$ and dispersion $V_\lambda$ for the $\lambda$-Gaussian channel with a given input distribution. Given the conditions that $(R_0-R_0^\prime)/\sigma>>1$ and $\Delta\leq\lambda \leq 1-\Delta$, where $\Delta>0$ is a given small value, we propose approximate calculations for $C_\lambda$ and $V_\lambda$. It is worth noting that the condition $(R_0-R_0^\prime)/\sigma>>1$ typically holds for the parameter settings provided in \cite{Ben,CZH,SongTcom2021,SongTcom2022}. The condition $\Delta\leq\lambda \leq 1-\Delta$ ensures that our analysis pertains to a non-trivial $\lambda$-Gaussian channel. 
	
	\subsubsection{Channel Capacity $C_\lambda$}
	Let $X\in\{0, 1\}$ and $Y$ be the input and output, respectively, of the $\lambda$-Gaussian channel, as illustrated in Fig.~\ref{fig:ESchannel}. The conditional PDF 
	of $Y$ given $X$ is expressed by 
	\begin{equation}
	p_{Y|X}(y|x)=(1-x)\Big(\lambda  \phi(y-R_0^\prime)+(1-\lambda )\phi(y-R_0)\Big)+x\phi(y- R_1)\nonumber.
	\end{equation}
	For input distribution Pr$(X=0)=1-q$ and Pr$(X=1)=q$, the marginal distribution of $Y$ is  
	\begin{eqnarray}
	p_Y(y)=\!\!\!\!\!\!\!\!\!\!&& (1-q)p_{Y|X}(y|0)+qp_{Y|X}(y|1)\nonumber\\
	=\!\!\!\!\!\!\!\!\!\!&&(1-q)\left(\lambda  \phi(y- R_0^\prime)+(1-\lambda )\phi(y- R_0)\right)+q\phi(y-R_1)\nonumber.
	\end{eqnarray}
	Thus, we have 
	\begin{align}
	\!\!I(X;Y)&=E\Bigg[\log\frac{P_{Y|X}(Y|X)}{P_{Y}(Y)}\Bigg]\nonumber\\
	= & qE\Bigg[\log\frac{\phi(Z)}{P_Y(Z+R_1)}\Bigg]+(1-q)\lambda E\Bigg[\log\frac{P_{Y|X}(Z+R_0^\prime|0)}{P_{Y}(Z+R_0^\prime)}\Bigg]\nonumber\\
	&\ \ \ \ \ \ \ \  +(1-q)(1-\lambda )E\Bigg[\log\frac{P_{Y|X}(Z+R_0|0)}{P_{Y}(Z+R_0)}\Bigg]\label{eq:Ixy}
	\end{align}
	where $Z~\sim\mathcal{N}(0, \sigma^2)$ is a Gaussian variable.
	Given the condition $(R_0-R_0^\prime)/\sigma>>1$ and $\Delta\leq\lambda \leq 1-\Delta$, we provide approximation calculation for each term of \eqref{eq:Ixy}. First, we have 
	\begin{align}
	& E\Bigg[\log P_{Y|X}(Z+R_0^\prime|0)\Bigg]\nonumber\\
	=  &\  E\Bigg[\log\Big(\lambda \phi(Z)+(1-\lambda)\phi(Z+R_0^\prime-R_0)\Big)\Bigg]\nonumber\\
	\leq &\ E\Bigg[\log\Big(\lambda \phi(Z)\Big)\Bigg]+\frac{1}{\Delta}e^{-\frac{\left(R_0-R_0^\prime-\beta\right)^2}{8\sigma^2}} \label{eq:Eapp}
	\end{align}
	where $\beta=\frac{2\sigma^2}{R_0-R_0^\prime}\log\frac{1-\Delta}{\Delta}$, which is small when $(R_0-R_0^\prime)/\sigma$ is large. 
	We provide the proof of \eqref{eq:Eapp} in Appendix~\ref{app:approx}. Therefore, the first term of \eqref{eq:Eapp} is a good approximation for $E[\log P_{Y|X}(Z+R_0^\prime|0)]$. 
	Similarly, we have 
	\begin{align}
	& E\Bigg[\log P_{Y|X}(Z+R_0|0)\Bigg]\approx E\Bigg[\log\Big((1-\lambda) \phi(Z)\Big)\Bigg] \nonumber\\
	& E\Bigg[\log P_{Y}(Z+R_1)\Bigg] \approx   E\Bigg[\log\Big((1-q)\lambda \phi(Z+R_1-R_0^\prime)+q\phi(Z)\Big)\Bigg]  \nonumber\\
	& E\Bigg[\log P_{Y}(Z+R_0^\prime)\Bigg] \approx   E\Bigg[\log\Big((1-q)\lambda \phi(Z)+q\phi(Z+R_0^\prime-R_1)\Big)\Bigg]  \nonumber\\
	& E\Bigg[\log P_{Y}(Z+R_0)\Bigg] \approx   E\Bigg[\log\Big((1-q)(1-\lambda) \phi(Z)\Big)\Bigg].  \nonumber
	\end{align}
	Note that the errors of these approximations are bounded by $\alpha e^{-{\left((R_0-R_0^\prime)/2-\beta\right)^2}/(2\sigma^2)}$, where $\alpha>0$. This can be strictly proven using the method presented in Appendix~\ref{app:approx}.
	
	After applying the approximations and performing some mathematical manipulations, we obtain 
	\begin{align}
	I(X;Y) \approx & H(q^\prime)-(1-q)H(\lambda )+q^\prime C_\text{BIAWGN}\Big(\frac{q}{q^\prime},\gamma\Big)\triangleq C_\lambda \label{eq:Clam} 
	\end{align}
	where $q^\prime=q+(1-q)\lambda , \gamma=\frac{R_0^\prime-R_1}{2\sigma}$, $H(x)=-x\log x-(1-x)\log(1-x)$ denotes the binary entropy function.
	$C_\text{BIAWGN}\big(q,\gamma\big)$
	represents the capacity of a BIAWGN channel with a Bernoulli $(q)$ input and a signal-to-noise (SNR) of $\gamma^2$, which is given by 
	\begin{align}
	C_\text{BIAWGN}\big(q,\gamma\big)=&-qE\Bigg[\log\Big(q+(1-q)e^{2\gamma(Z-\gamma)}\Big)\Bigg] \nonumber\\
	&\ \ \ -(1-q)E\Bigg[\log\Big(qe^{2\gamma(Z-\gamma)}+1-q\Big)\Bigg].\nonumber
	\end{align} 
	
	Interestingly, when $\lambda=0$, we have $C_\lambda=H(q)$, which is the capacity of a noiseless channel. This can be interpreted as that when $\lambda=0$ and $(R_0-R_0^\prime)/\sigma>>1$ leading to $(R_0-R_1)/\sigma>>1$, the channel becomes highly reliable. On the other hand, when $\lambda=1$, we have $C_\lambda=C_\text{BIAWGN}\big(q,\gamma\big)$, which corresponds to the capacity of the BIAWGN channel with an SNR of $\gamma^2$. Therefore, the expression of $C_\lambda$ remains valid in these two extreme cases. 
	
	It is worth mentioning that we can obtain an alternative form of $C_\lambda$ using the following method. We define a Boolean variable $V\in \{0, 1\}$ as an indicator to determine whether an HRS occurs. Specifically, $V=1$ if $Y=R_0+Z$ and $V=0$ otherwise.
	Thus, we have $\text{Pr}(V=0)=q^\prime$ and $\text{Pr}(V=1)=1-q^\prime$. When $V=0$, $Y$ takes the form of $Y=R_1+Z$ with probability $q/q^\prime$ or $Y=R_0^\prime+Z$  with probability $1-q/q^\prime$. Furthermore, we can demonstrate that $H(Y)\approx H(Y,V)$, which subsequently implies $H(X|Y)\approx H(X|Y,V)$, using a technique similar to that employed in \eqref{eq:Eapp}.
	The mutual information is then derived as
	\begin{align}
	I(X;Y)& =H(X)-H(X|Y)\nonumber\\
	&\approx H(X)-H(X|Y,V)\nonumber\\
	&=H(X)-(1-q^\prime)H(X|Y, V=1)-q^\prime H(X|Y, V=0)\nonumber\\
	&=H(X)-q^\prime \big(H(X|V=0)-I(X;Y|V=0)\big)\nonumber\\
	&=H(q)-q^\prime H\Big(\frac{q}{q^\prime}\Big)+q^\prime C_\text{BIAWGN}\Big(\frac{q}{q^\prime},\gamma\Big)\label{eq:alterC}.
	\end{align}
	In \eqref{eq:alterC} we have used the fact that given $V=0$, $X$ is a Bernoulli variable with $\text{Pr}(X=1)=q/q^\prime$ and $\text{Pr}(X=0)=1-q/q^\prime$. Additionally, given $V=0$,  the pair $(X, Y)$ is an input and output of a BIAWGN channel. It can be verified that \eqref{eq:alterC} and \eqref{eq:Clam}  are equivalent. 
	

	\subsubsection{Channel Dispersion $V_{\lambda}$}
	The channel dispersion is defined as the variance of $\log\frac{P_{Y|X}(Y|X)}{P_{Y}(Y)}$.  We utilize approximations similar to those presented in the previous sub-section, which are expressed in the form 
	\begin{align}
	E\Bigg[\Bigg(\log P_{Y|X}(Z+R_0^\prime|0)\Bigg)^2\Bigg]\approx &\ E\Bigg[\Bigg(\log\Big(\lambda \phi(Z)\Big)\Bigg)^2\Bigg]\label{eq:approx2}.
	\end{align}
	The validity of this approximation is rigorously proven in Appendix~\ref{app:approx2}. It demonstrate that, for $(R_0-R_0^\prime)/\sigma>>1$ and $\Delta\leq\lambda \leq 1-\Delta$, the error of this approximation is bounded by 
	$\Big(\frac{3(R_0-R_0^\prime)^4}{4\sigma^4}+\frac{3}{2\Delta^2}\Big)e^{-\frac{(R_0-R_0^\prime-\beta)^2}{8\sigma^2}}$, which is very small when $(R_0-R_0^\prime)/\sigma$ is large. 
	Thus, we have 
	\begin{align}
	V_{\lambda}  =&\ E\Bigg[\Bigg(\log\frac{P_{Y|X}(Y|X)}{P_{Y}(Y)}\Bigg)^2\Bigg]- I(X;Y)^2\nonumber\\
	\approx &\ qE\Bigg[\Bigg(\log\frac{\phi(Z)}{(1-q)\lambda \phi(Z+R_1-R_0^\prime)+q\phi(Z)}\Bigg)^2\Bigg]\nonumber\\
	&\ \  +(1-q)\lambda E\Bigg[\Bigg(\log\frac{\lambda \phi(Z)}{(1-q)\lambda \phi(Z)+q\phi(Z+R_0^\prime-R_1)}\Bigg)^2\Bigg]\nonumber\\
	&\ \ \ \ \ \ \ \  +(1-q)(1-\lambda )\big(\log(1-q)\big)^2-(C_\lambda)^2\nonumber\\
	=\ & qE\Bigg[\Bigg(\log\Big((1-q)\lambda e^{2\gamma(\widetilde{Z}-\gamma)}+q\Big)\Bigg)^2\Bigg]\nonumber\\
	&\ \  +(1-q)\lambda E\Bigg[\Bigg(\log\Big(1-q+\frac{q}{\lambda} e^{2\gamma(\widetilde{Z}-\gamma)}\Big)\Bigg)^2\Bigg]\nonumber\\
	&\ \ \ \ \ \ \ \  +(1-q)(1-\lambda )\big(\log(1-q)\big)^2-(C_\lambda)^2\label{eq:Vlam}
	\end{align}
	where $\widetilde{Z}\sim \mathcal{N}(0,1)$ is a Gaussian variable with mean 0 and unitary variance. Therefore, the expression of $C_\lambda$ remains valid in  these two extreme cases. Note that \eqref{eq:Vlam} remains valid when $\lambda=1$ and $\lambda\rightarrow 0$.
	
			\begin{figure*}[t]
		\includegraphics[width=
		6.3 in]{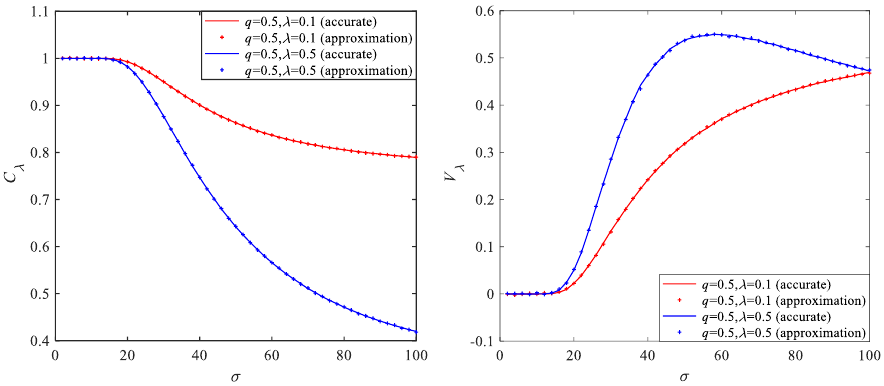}
		\centering
		\caption{Channel capacity and dispersion of $\lambda$-Gaussian channel: accurate values and approximations. The associated resistance values are $R_0=1000\ \Omega, R_1=100\ \Omega$,  and $R_s=250\ \Omega$.} \label{fig:capacity}
	\end{figure*}

	Fig.~\ref{fig:capacity} illustrates both the accurate and approximate channel capacity $C_\lambda$ and dispersion $V_\lambda$ of $\lambda$-Gaussian channel. With typical parameter settings of $R_0=1000\ \Omega, R_1=100\ \Omega$,  and $R_s=250\ \Omega$,  we observe almost no difference between the accurate and approximate values.

	\section{Density Evolution and Iterative Decoding Threshold Over $\lambda$-Gaussian Channel}\label{sec:threshold}
	
	In this section, we propose a DE framework for the deign of practical sparse-graph codes over a $\lambda$-Gaussian channel. Sparse-graph codes, including LDPC codes and repeat-accumulate (RA) codes, have been proven to be asymptotic capacity-approaching \cite{LinBook}. A prominent feature of these codes is that their decoding can be efficiently implemented on a sparse factor graph using the belief propagation (BP) algorithm. By leveraging the iterative nature of BP, DE is commonly employed for predicting decoding performance and optimizing the codes. By tracking the PDF of the log-likelihood ratio (LLR) message, a decoding threshold can be determined under the assumption of infinite code length. 
	We demonstrate the Gaussanity of the LLR  messages over the $\lambda$-Gaussian channel. Based on this, we develop a low-complexity two-dimensional DE framework for code optimization.  The analysis applies to a general sparse-graph code.

	\subsection{Gaussian Approximation of LLR}
	In this section, we demonstrate that the initial LLR provided to the decoder, as derived in \eqref{eq:LLR}, can be approximated as Gaussian. Furthermore, when $\lambda\approx 0$ and $1$, the channel can be well approximated by a noiseless channel and a BIAWGN channel with an SNR of $\gamma^2$, respectively. In these cases, the LLR exhibits a Gaussian distribution. Therefore, our subsequent analysis will concentrate on the scenario where $\Delta\leq\lambda \leq 1-\Delta$, demonstrating the Gaussian nature of the LLR. Since the $\lambda$-Gaussian channel produce three distinct types of outputs, specifically $Y=r+Z$, where $r$ can be $R_0, R_1$, and $R_0^\prime$, corresponding to the readback signals from the HRS, LRS, and SP cells of ReRAM, respectively, the LLR exhibits three different PDFs. We will individually establish the Gaussian nature of these PDFs.
	
	\subsubsection{HRS Cell} For $Y=R_0+Z$ with $Z\sim\mathcal{N}(0,\sigma^2)$, the channel input should be $X=0$. In this scenario, $Y$ can be considered as the readback signal of an HRS cell in ReRAM. The LLR message \eqref{eq:LLR} can be reformulated as
	\begin{eqnarray}
	L_\text{HRS}=&&\!\!\!\!\!\!\!\!\log\frac{(1-q)(1-\lambda)\phi(Y, R_0)}{q\phi(Y, R_1)}+\log\left(1+\frac{\lambda \phi(Y, R_0^\prime)}{(1-\lambda)\phi(Y, R_0)}\right)\nonumber\\
	=&&\!\!\!\!\!\!\!\! \frac{(R_0-R_1)(R_0-R_1+2Z)}{2\sigma^2}-\log\frac{q}{(1-q)(1-\lambda)}\nonumber\\
	&&\!\!\!\!\!\!\!\!\ \ \ \ \ \ \ \ +\log\left(1+\frac{\lambda}{1-\lambda}e^{-\frac{(R_0-R_0^\prime)(R_0-R_0^\prime+2Z)}{2\sigma^2}}\right)\label{eq:neglect1}\\
	\approx &&\!\!\!\!\!\!\!\!\frac{(R_0-R_1)(R_0-R_1+2Z)}{2\sigma^2}-\log\frac{q}{(1-q)(1-\lambda)}\nonumber
	\end{eqnarray}
	where we have neglected the last term of \eqref{eq:neglect1} with the assumption of $(R_0-R_0^\prime)/\sigma>>1$. As shown in Appendix~\ref{app:approx}, the expected value of the absolute error of this approximation is in the form of $\alpha e^{-\frac{(R_0-R_0^\prime-\beta)^2}{8\sigma^2}}$ where $\alpha>0$ and $\beta=\frac{2\sigma^2}{R_0-R_0^\prime}\log\frac{1-\Delta}{\Delta}$ for $\Delta<\lambda <1-\Delta$. Therefore, the LLR can be approximated as Gaussian  
	\begin{equation}
	L_\text{HRS}\sim\mathcal{N}\left(\frac{(R_0-R_1)^2}{2\sigma^2}-\log\frac{q}{(1-q)(1-\lambda)}, \frac{(R_0-R_1)^2}{\sigma^2}\right)\nonumber .
	\end{equation}
	Since $(R_0-R_0^\prime)/\sigma>>1$, the mean value of $L_\text{HRS}$ is considerably large. Therefore, the readback signal of an HRS cell is quite reliable. During the DE process, we assume the message associated with an HRS cell is reliable.      
	\subsubsection{LRS Cell} Similarly, for $Y=R_1+Z$, the channel input should be $X=1$. In this scenario, $Y$ can be considered as the readback signal of an LRS cell in ReRAM. The LLR message \eqref{eq:LLR} can be reformulated as
	\begin{eqnarray}
	L_\text{LRS}=&&\!\!\!\!\!\!\!\!\log\frac{\lambda(1-q)\phi(Y, R_0^\prime)}{q\phi(Y, R_1)}+\log\left(1+\frac{(1-\lambda) \phi(Y, R_0)}{\lambda\phi(Y, R_0^\prime)}\right)\nonumber\\
	=&&\!\!\!\!\!\!\!\! -\frac{(R_0^\prime-R_1)(R_0^\prime-R_1-2Z)}{2\sigma^2}-\log\frac{q}{\lambda(1-q)}\nonumber\\
	&&\!\!\!\!\!\!\!\!\!\!\ \ \ \ \ \ \ \ +\log\left(1+\frac{1-\lambda}{\lambda}e^{-\frac{(R_0-R_0^\prime)(R_0+R_0^\prime-2R_1-2Z)}{2\sigma^2}}\right)\label{eq:neglect}\\
	\approx &&\!\!\!\!\!\!\!\!-\frac{(R_0^\prime-R_1)(R_0^\prime-R_1-2Z)}{2\sigma^2}-\log\frac{q}{\lambda(1-q)}\nonumber.
	\end{eqnarray}
	Therefore, the LLR can be approximated as Gaussian 
	\begin{equation}
	L_\text{LRS}\sim\mathcal{N}\left(-\frac{(R_0^\prime-R_1)^2}{2\sigma^2}-\log\frac{q}{\lambda(1-q)}, \frac{(R_0^\prime-R_1)^2}{\sigma^2}\right)\nonumber .
	\end{equation}
	
			\begin{figure*}[t]
		\includegraphics[width=
		6.2 in]{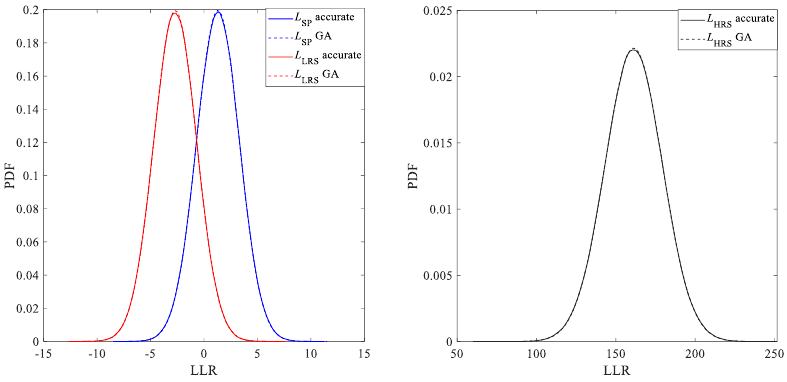}
		\centering
		\caption{Accurate PDFs of $L_\text{LRS}, L_\text{SP}, L_\text{HRS}$ and their Gaussian approximation (GA) for $\lambda=1/2$ and $q=1/2$. The associated resistance values are $R_0=1000\ \Omega, R_1=100\ \Omega$,  and $R_s=250\ \Omega$, and the noise  standard deviation is $\sigma=50$.} \label{fig:GaussAPP}
	\end{figure*}

	\subsubsection{SP Cell} For $Y=R_0^\prime+Z$, the channel input should be $X=0$. In this scenario, $Y$ can be considered as the readback signal of an SP cell in ReRAM. We follow a similar procedure and obtain
	\begin{equation}
	L_\text{SP}
	\approx \frac{(R_0^\prime-R_1)(R_0^\prime-R_1+2Z)}{2\sigma^2}-\log\frac{q}{\lambda(1-q)}\nonumber.
	\end{equation}
	Thus, we have  
	\begin{equation}
	L_\text{SP}\sim\mathcal{N}\left(\frac{(R_0^\prime-R_1)^2}{2\sigma^2}-\log\frac{q}{\lambda(1-q)}, \frac{(R_0^\prime-R_1)^2}{\sigma^2}\right)\nonumber .
	\end{equation}
	For all the three cases, the commonly used assumption that the variance of LLR is equal to twice its mean is also approximately satisfied.

	We compared the PDFs of $L_\text{HRS}, L_\text{LRS}$ and $L_\text{SP}$ with their Gaussian approximation in Fig.~\ref{fig:GaussAPP} for $\lambda=1/2$ and $q=1/2$. It is evident that the Gaussian approximation is quite accurate.  Furthermore, with typical parameter settings of $R_0=1000\ \Omega, R_1=100\ \Omega$, and $R_s=250\ \Omega$, $L_\text{HRS}$ remains quite reliable even at a high noise level of $\sigma=50$.

	
	\subsection{Two-Dimensional Representation of Message PDF}
	
	Due to the message asymmetry, assuming the LLRs to be single-parameter Gaussian is unreasonable. Based on the analysis presented in the previous sub-section, it is safe to assume the magnitude of $L_\text{HRS}$ as infinite, i.e.,  the initial LLR of the HRS cell is reliable. Since the probability of $r=R_0$ is
	\begin{equation}
	p_0=(1-q)(1-\lambda)\nonumber
	\end{equation}
	with probability $p_0$, the initial message to the decoder is reliable. 
	The reliable message serves as side information during iterative decoding, and does not require updating.  
	
	LLRs of an LRS and an SP cells exhibit similar magnitudes, differing only by the term $\pm\log\frac{q}{\lambda(1-q)}$. To see the magnitude of the LLR of $L_\text{LRS}$, we should normalize it as $-L_\text{LRS}$, since the message is associated with data $X=1$. 
	We approximate the LLRs of both LRS and SP cells using a single Gaussian variable with mean of  
	\begin{equation}
	m_0=\frac{(R_0^\prime-R_1)^2}{2\sigma^2}+\frac{q-\lambda(1-q)}{q+\lambda(1-q)}\log\frac{q}{\lambda(1-q)} \nonumber.
	\end{equation}
	This mean $m_0$ is a weighted average of the means of $-L_\text{LRS}$ and $L_\text{SP}$, with weights corresponding to their fractions $\frac{q}{q+\lambda(1-q)}$ and $\frac{\lambda(1-q)}{q+\lambda(1-q)}$.
	Specifically, we assume that the initial LLRs of both LRS and SP cells follow a Gaussian distribution $\mathcal{N}\left(m_0, 2m_0\right)$.

	Based on the above analysis, we utilize the tuple $(p, m)$ to specify the PDF of a message employed in decoding over a $\lambda$-Gaussian channel.  Specifically, $p$ represents the probability that the associated coded bit has already been recovered. Conversely, with probability $1-p$, the coded bit remains unrecovered, and its LLR follows a Gaussian distribution of $\mathcal{N}(m, 2m)$. Thus, the tuple $(p, m)$ fully captures the statistical characteristics of the message.

	\subsection{Density Evolution and Decoding Threshold}
	
\begin{figure}[t]
		\includegraphics[width=
		2.2 in]{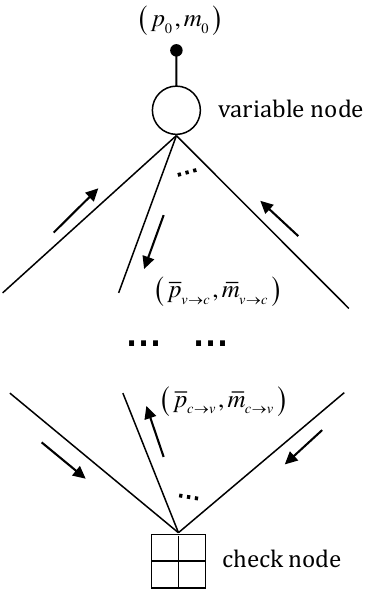}
		\centering
		\caption{Message passing over factor graph.} \label{fig:tanner}
	\end{figure}

	The decoder employs a standard BP decoding process on a sparse factor graph to recover the data. The graph comprises two types of nodes: variable and check nodes. The decoding is accomplished through local maximum a posteriori (MAP) processing conducted by both the variable and check nodes, facilitated by the propagation of messages between them. Fig.~\ref{fig:tanner} presents a general factor graph along with a schematic diagram illustrating the message passing.
	To track the PDF tuple of the message, we explore the update rules of both variable and check nodes. These rules are  
	used to analyze the iterative decoding error probability and determine the decoding threshold. Our analysis is directed towards a general factor graph, thereby rendering it applicable to a wide range of sparse-graph codes.  
	
	\subsubsection{Variable Node Updating}
	
	During iterative decoding, a variable node updates its output based on the incoming message from the check nodes in its neighborhood. 
	According to the standard BP, a degree-$d$ variable node updates its message based on $d-1$ LLR messages from check nodes and one initial LLR from the channel receive. The processing is a MAP estimation 
	of a variable based on $d$ independent observations. Specifically, let $(\bar{p}_{c\rightarrow v}, \bar{m}_{c\rightarrow v})$ be the PDF tuple for message from the check nodes. As presented in the previous sub-section, the PDF tuple of message from the $\lambda$-Gaussian channel is $(p_0, m_0)$. Thus, the PDF tuple $(p_v, m_v)$ of
	a degree-$d$ variable node output is derived as
	\begin{eqnarray}
	p_v=&&\!\!\!\!\!\!\!\!\!\! 1-(1-\bar{p}_{c\rightarrow v})^{d-1}(1-p_0)\label{eq:pv}\\
	m_v =&&\!\!\!\!\!\!\!\!\!\! m_0+(d-1)\bar{m}_{c\rightarrow v}\nonumber. 
	\end{eqnarray}
	Here, \eqref{eq:pv} is derived based on the fact that the output of a variable node is reliable if and only if at least one of its inputs is reliable. Let  $\{a_d, d=1, 2,..., d_v\}$ represent the degree distribution of variable nodes, where $a_d$ denotes the proportion of edges incident to variable nodes of
	degree $d$. The message outputted from the variable nodes after being averaged concerning its degree distribution is
	\begin{eqnarray}
	\bar{p}_{v\rightarrow c}=&&\!\!\!\!\!\!\!\!\!\! 1-(1-p_0)\sum_{d=1}^{d_v}a_d(1-\bar{p}_{c\rightarrow v})^{d-1}\nonumber\\
	\bar{m}_{v\rightarrow c} =&&\!\!\!\!\!\!\!\!\!\! m_0+\left(\sum_{d=1}^{d_v}a_d d-1\right)\bar{m}_{c\rightarrow v}\nonumber. 
	\end{eqnarray}
	
	\subsubsection{Check Node Updating}
	A degree-$d$ check node updates its output based on $d-1$ incoming LLR messages from the variable nodes in its neighborhood. 
	This operation follows the same rule as a MAP decoding  
	of a single parity check code.  Since a check node will output a reliable message if and only if all of its $d-1$ incoming messages are reliable, we have 
	\begin{equation}
	p_c= (\bar{p}_{v\rightarrow c})^{d-1}\nonumber.
	\end{equation}
	If $i$ (where $0<i\leq d-1$) incoming messages are unreliable, the output will be unreliable. In this case, according to the check node update rule \cite{GA}, 
	the mean of its output is 
	\begin{equation}
	f(i)\triangleq\Phi^{-1}\left(1-\left(1-\Phi(\bar{m}_{v\rightarrow c})^i\right)\right)
	\end{equation}
	where 
	\begin{equation} 
	\Phi(x)= \left\{ \begin{aligned} & 1-\frac{1}{\sqrt{4\pi x}}\int\tanh\left(\frac{u}{2}\right)e^{-\frac{(u-x)^2}{4x}}du\ \  x>0 \\ &  1 \ \ \ \ \ \ \ \ \ \ \ \ \ \ \  \ \ \ \ \ \ \ \ \ \ \ \ \ \ \ \ \ \ \ \  \ \ x=0 \end{aligned} \right. 
	\end{equation}
	Since $i$ follows a binomial distribution, we can obtain the mean output of the check node as
	\begin{equation}
	m_c = \sum_{i=1}^{d-1}\binom{d-1}{i}(\bar{p}_{v\rightarrow c})^{d-1-i}(1-\bar{p}_{v\rightarrow c})^if(i)\nonumber .
	\end{equation}
	By averaging the check node output $(p_c, m_c)$ concerning the degree distribution $\{b_d, d=1, 2,..., d_c\}$, we obtain
	\begin{eqnarray}
	\bar{p}_{c\rightarrow v}=&&\!\!\!\!\!\!\!\!\!\! \sum_{d=1}^{d_c}b_d(\bar{p}_{v\rightarrow c})^{d-1}\nonumber\\
	\bar{m}_{c\rightarrow v} =&&\!\!\!\!\!\!\!\!\!\! \sum_{d=1}^{d_c}b_d \sum_{i=1}^{d-1}\binom{d-1}{i}(\bar{p}_{v\rightarrow c})^{d-1-i}(1-\bar{p}_{v\rightarrow c})^if(i)\nonumber. 
	\end{eqnarray}
	Here, $b_d$ represent the fraction of edges incident to check nodes of
	degree $d$.
	
	\subsubsection{Decoding Threshold $\sigma_\text{th}$}
	After a sufficient number of iterations, the two-dimensional PDF of LLR message outputted one each edge will converge to a fixed point, denoted as $(p_{\infty},m_{\infty})$. Thus, the error 
	probability after a hard decision is derived as
	\begin{equation}
	P_e=(1-p_{\infty})Q\left(\sqrt{\frac{m_{\infty}}{2}}\right)\nonumber.
	\end{equation}
	Note that, given $\lambda$ and the degree distributions $\{a_d\}$ and $\{b_d\}$, $P_e$ is solely determined by the channel noise parameter $\sigma$. The decoding threshold, denoted by $\sigma_\text{th}$, is the maximum tolerable value of $\sigma$ that makes $P_e=0$. 
	
\begin{figure}[t]
	\includegraphics[width=
	3.6 in]{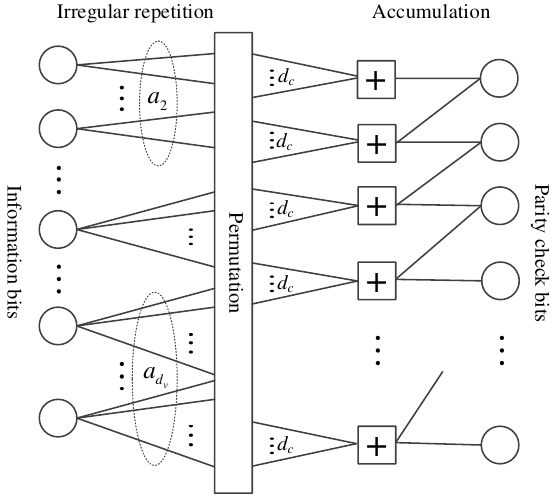}
	\centering
	\caption{Factor graph of IRA code.} \label{fig:IRA}
\end{figure}

	\section{Code Optimization and Numerical Results}\label{sec:numerical}
	This section illustrates how to design a channel code utilizing our channel decomposition theory. 
	Although our method applies to the design of a general sparse-graph code,
	we employ irregular repeat-accumulate (IRA) code as an illustrative example to demonstrate our results. The IRA code is an asymptotic capacity-approaching sparse-graph code with low encoding and decoding complexities \cite{IRA}. Fig.~\ref{fig:IRA} depicts the factor graph of this code. The IRA code employed here is systematic, featuring an irregular variable node degree and a constant check node degree. Our objective is to optimize the degree distribution, and within our simulation, the permutation used in the IRA encoder is randomly generated.
	
	Suppose we design codes for ReRAM arrays with size $N\times N$, where $N=128$. To ensure that the code can tolerate up to $K$ SFs, we will design the code under the $\lambda^*$-Gaussian channel, where $\lambda^*=E[\lambda_N]+3\sqrt{D[\lambda_N]}$. This design approach enables the code to address the SP rates for $\text{Pr}(\lambda_N<\lambda^*)\approx 99.87\%$ of the data patterns. The code design leverages our proposed two-dimensional DE framework, detailed in Sec.~\ref{sec:threshold}. The code optimization follows the following procedure. We utilize standard parameter values of $R_0=1000\ \Omega, R_1=100\ \Omega$, and $R_s=250\ \Omega$. We consider constant check node degree of $d_c$ and 3 types of variable node degrees of 3, 10, and 36. Then we search the optimal variable node degree distribution $\{a_3, a_{10}, a_{36}\}$ that achieves the target code rate $R$ and also possesses the optimal decoding threshold $\sigma_\text{th}$.  Following this procedure, we have obtained the IRA codes listed in TABLE~\ref{tab:code} that perform very close to the Shannon limit $\sigma^*$. Here $\sigma^*$ is the maximum tolerable noise level that supports the information rate $R$ required by channel capacity $C_{\lambda^*}$.  
	
	\begin{table}
		\caption{Rate-$R$ IRA codes tailored for $128\times128$ ReRAM arrays with $K$ SFs in the arrays. The codes are designed under the $\lambda^*$-Gaussian channel. $\sigma_\text{th}$ is the decoding threshold, and $\sigma^*$ is the Shannon limit.}
		\label{tab:code}
		\begin{center}
			\begin{tabular}{|c|c|c|c c c|c|c|}
				\hline\hline
				\multicolumn{8}{|c|}{$R=0.5$}\\
				\hline
				$K$&$\lambda^*$&$d_c$&$a_3$&$a_{10}$&$a_{36}$ & $\sigma_\text{th}$&$\sigma^*$ \\
				\hline
				$2$&$0.5338$&$6$&$0.3561$&$0.4165$&$0.2274$& $65$&	$66$\\
				\hline
				$5$&$0.8306$&$6$&$0.3704$&$0.3560$&$0.2736$&$50$&$52$\\
				\hline\hline
				\multicolumn{8}{|c|}{$R=0.8$}\\
				\hline
				$K$&$\lambda^*$&$d_c$&$a_3$&$a_{10}$&$a_{36}$ & $\sigma_\text{th}$&$\sigma^*$ \\
				\hline
				$1$&$0.3398$&$16$&$0.6540$&$0.3100$&   $0.0360$& $37$&	$39$\\
				\hline
				$2$&$0.5338$&$16$&$0.6878$&$0.1670$& $0.1452$&$33$&$35$\\
				\hline\hline
			\end{tabular}
		\end{center}
	\end{table} 

\begin{figure}[t]
	\includegraphics[width=
	4.4 in]{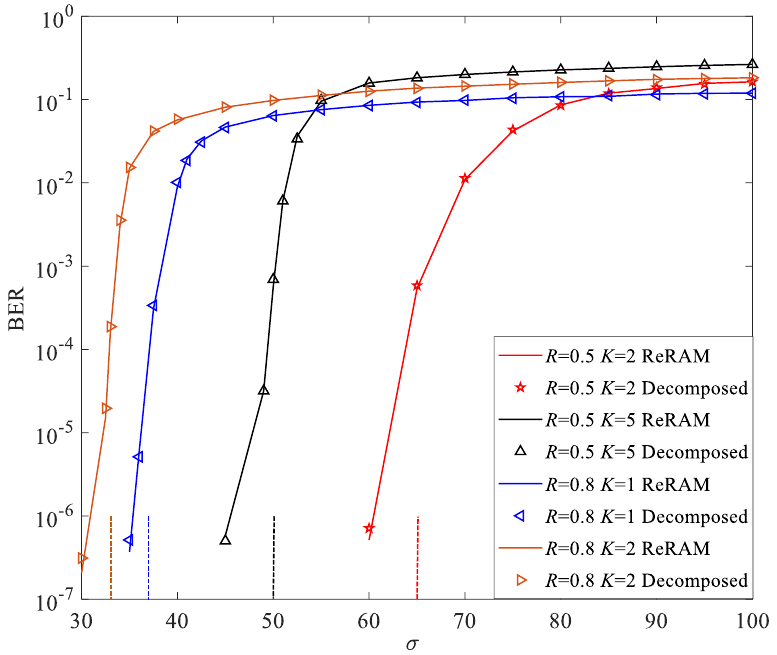}
	\centering
	\caption{BER of the codes listed in TABLE~\ref{tab:code} over both ReRAM and the decomposed channels. The dashed lines illustrate the corresponding decoding thresholds over the $\lambda^*$-Gaussian channel.} \label{fig:BER}
\end{figure}

\begin{figure}[t]
	\includegraphics[width=
	4.4 in]{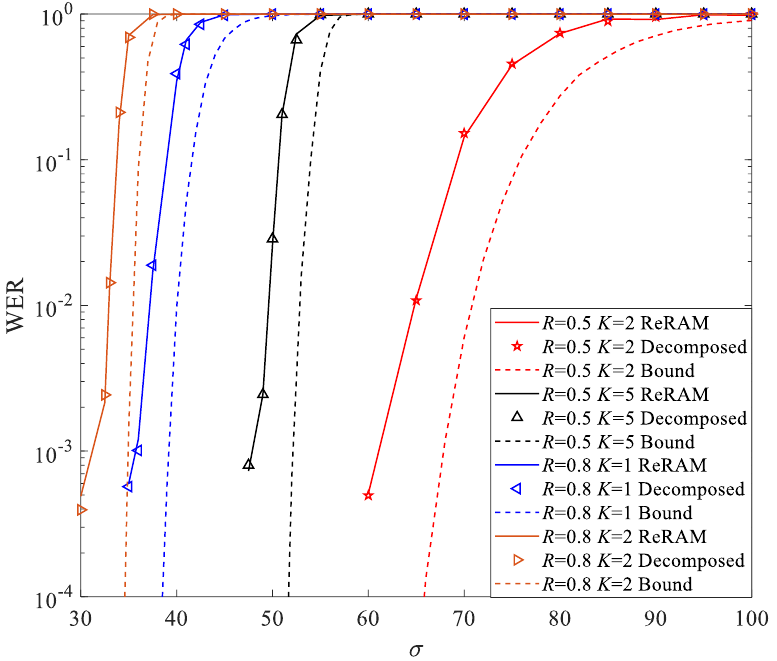}
	\centering
	\caption{WER of the codes listed in TABLE~\ref{tab:code} over both ReRAM and the decomposed channels.} \label{fig:WER}
\end{figure}
	
	Simulation results for the bit error rate (BER) and word error rate (WER)  of the codes listed in TABLE~\ref{tab:code} are illustrated in Figs.~\ref{fig:BER} and \ref{fig:WER}, respectively. These codes are simulated over ReRAM channel configured with memory array size of $N\times N$, where $N=128$. Note that codes designed targeting to $K$ SFs can still work when the actual number of SFs in the array is fewer than $K$. However, to compare the simulation results with the corresponding decoding thresholds, we have simulated the code for ReRAM with exactly $K$ active SFs.  In each simulation, the locations of the SFs within the array are randomized. The standard BP decoding, as previously specified, is employed, with the channel parameter $\lambda$ being estimated using \eqref{eq:channel_est} in each run. To validate our channel decomposition approach, we have also conducted simulations under the corresponding decomposed channel model, which is a mixed $\lambda$-Gaussian channel characterized by a mixing proportion of $F_N(\lambda)$.
	
	In both figures, the simulation results obtained over the ReRAM channel coincide with those over the mixed $\lambda$-Gaussian channel, thereby confirming the validity of channel decomposition approach. In Fig.~\ref{fig:BER}, the gaps between the BER curves (at $10^{-6}$) and the decoding thresholds $\sigma_\text{th}$ over the corresponding $\lambda^*$-Gaussian channel are relatively small. 
	In Fig.~\ref{fig:WER}, we compared the WERs of our code with finite-length performance bound given by \eqref{eq:bound}. It appears that there is a narrow gap between the WER curves and their respective performance bounds. However, an error floor is observed in the region where BER drops below $10^{-6}$ and the WER falls below $10^{-3}$. One plausible explanation for this phenomenon is the random generation of the permutation employed in the IRA encoder. Strategies for constructing permutations with lower error floors can be found in \cite{LinBook}.

	\section{Conclusion} \label{sec:end}
		We proposed a channel decomposition approach for performance analysis and channel code design over ReRAM channels. This approach decomposes the non-ergodic and data-dependent ReRAM channel into multiple stationary memoryless channels. By analyzing the capacity and dispersion of these stationary memoryless channels, we derived a finite-length performance bound. Leveraging this decomposition, we proposed a practical sparse-graph code design that utilizes the newly developed two-dimensional DE. Taking IRA code as an illustrative example, we demonstrated our numerical results for code design. The designed codes not only exhibit capacity-approaching decoding thresholds but also yield BER simulation results that are close to the derived finite-length performance bound. Furthermore, simulations were conducted to confirm the equivalence between the original ReRAM channel and the decomposed channels.

	\appendix
	
	\subsection{Proof of \eqref{eq:Eapp}} \label{app:approx}
	
	Denote $\Delta R=R_0-R_0^\prime$. For $\lambda \geq \Delta>0$, we have
	\begin{align}
	&  E\Bigg[\log\Big(\lambda  \phi(Z)+(1-\lambda )\phi(Z+R_0^\prime-R_0)\Big)\Bigg]\nonumber\\
	=&\ E\Bigg[\log\Big(\lambda  \phi(Z)\Bigg]+E\Bigg[\log\Bigg(1+\frac{1-\lambda }{\lambda }e^{-\frac{\Delta R(\Delta R-2Z)}{2\sigma^2}}\Bigg)\Bigg]\nonumber
	\end{align}
	where $Z~\sim\mathcal{N}(0, \sigma^2)$.
	Thus, we only need to show that the second term of the above expression is upper bounded by $\frac{1}{\Delta}e^{-\frac{(\Delta R-\beta)^2}{8\sigma^2}}$ where $\beta=\frac{2\sigma^2}{\Delta R}\log\frac{1-\Delta}{\Delta}$.  	
	
	Let $\tau=\frac{\Delta R}{2}+\frac{\sigma^2}{\Delta R}\log\frac{\lambda}{1-\lambda}$. Since $\Delta R/\sigma>>1$, it is safe to assume $\Delta R>\tau>0$. We have 
	\begin{equation}
	\frac{1-\lambda }{\lambda }e^{-\frac{\Delta R(\Delta R-2Z)}{2\sigma^2}}> 1 \iff Z>\tau.\nonumber
	\end{equation}
	Utilizing the inequalities
	\begin{align}
	\log(1+x)&\leq x \   \  \text{for}\  x> 0 \nonumber\\
	\log (1+x)&\leq 1+\log x \ \ \text{for}\  x> 1 \nonumber
	\end{align}
	we obtain
	\begin{align}
	E\Bigg[\log\Bigg(&1+\frac{1-\lambda }{\lambda }e^{-\frac{\Delta R(\Delta R-2Z)}{2\sigma^2}}\Bigg)\Bigg]\leq \ E\Bigg[\frac{1-\lambda }{\lambda }e^{-\frac{\Delta R(\Delta R-2Z)}{2\sigma^2}}1\Big\{Z\leq\tau\Big\}\Bigg]\nonumber \\
	&   +E\Bigg[\Big(1+\log \frac{1-\lambda }{\lambda }-\frac{\Delta R(\Delta R-2Z)}{2\sigma^2}\Big)1\Big\{Z>\tau\Big\}\Bigg]\label{eq:Evv}.
	\end{align}
	The first two term of the above expression is bounded by
	\begin{align}
	E\Bigg[\frac{1-\lambda }{\lambda }e^{-\frac{\Delta R(\Delta R-2Z)}{2\sigma^2}}1\Big\{Z\leq\tau\Big\}\Bigg]&= \frac{1-\lambda }{\lambda } Q\Big(\frac{\Delta R-\tau}{\sigma}\Big)\nonumber\\
	&\leq \frac{1-\lambda }{2\lambda }e^{-\frac{(\Delta R-\tau)^2}{2\sigma^2}}\label{eq:firstE}.
	\end{align}
	Using the formulations
	\begin{align} 
	E\Big[1\{Z>\tau\}\Big]&=Q\big(\frac{\tau}{\sigma}\big)\leq\frac{1}{2}e^{-\frac{\tau^2}{2\sigma^2}} \label{eq:Q0}\\
	E\Big[1\{Z>\tau\}Z\Big]&=\frac{\sigma}{\sqrt{2\pi}}e^{-\frac{\tau^2}{2\sigma^2}}\nonumber
	\end{align}
	the second  term is upper bounded by
	\begin{align}
	&\Bigg(\frac{1}{2}+\frac{1}{2}\log\frac{1-\lambda }{\lambda }-\frac{\Delta R^2}{4\sigma^2}+\frac{\Delta R}{\sigma\sqrt{2\pi}} \Bigg)e^{-\frac{\tau^2}{2\sigma^2}}\nonumber\\
	\leq \ &\Bigg(\frac{1}{2}+\frac{1}{2}\log\frac{1-\lambda }{\lambda }+\frac{1}{2\pi}\Bigg)e^{-\frac{\tau^2}{2\sigma^2}} 
	\leq \frac{1+\lambda }{2\lambda }e^{-\frac{\tau^2}{2\sigma^2}} \label{eq:secondE2}.
	\end{align}
	where we used $\log x<x$ and $\frac{1}{2\pi}<1/2$ . Applying \eqref{eq:firstE} and \eqref{eq:secondE2} to \eqref{eq:Evv}, and using $\lambda>\Delta$ and $\min\{\Delta R-\tau, \tau\}\geq \frac{\Delta R}{2}-\frac{\sigma^2}{\Delta R}\log\frac{1-\Delta}{\Delta}$, we obtain 
	\begin{align}
	E\Bigg[\log\Bigg(1+\frac{1-\lambda }{\lambda }e^{-\frac{\Delta R(\Delta R-2Z)}{2\sigma^2}}\Bigg)\Bigg]&\leq \frac{1}{\Delta}e^{-\frac{\big(\frac{\Delta R}{2}-\frac{\sigma^2}{\Delta R}\log\frac{1-\Delta}{\Delta}\big)^2}{2\sigma^2}} =\frac{1}{\Delta}e^{-\frac{(\Delta R-\beta)^2}{8\sigma^2}}\nonumber.
	\end{align}
	
	\subsection{Proof of \eqref{eq:approx2}} \label{app:approx2}
	
	Denote $\Delta R=R_0-R_0^\prime$. For $\lambda \geq \Delta>0$, we have
	\begin{align}
	&  E\Bigg[\Bigg(\log\Big(\lambda  \phi(Z)+(1-\lambda )\phi(Z+R_0^\prime-R_0)\Big)\Bigg)^2\Bigg]\nonumber\\
	\leq &\ E\Bigg[\Bigg(\log\Big(\lambda  \phi(Z)\Bigg)^2\Bigg]+E\Bigg[\Bigg(\log\Bigg(1+\frac{1-\lambda }{\lambda }e^{-\frac{\Delta R(\Delta R-2Z)}{2\sigma^2}}\Bigg)\Bigg)^2\Bigg]\label{eq:Efai3}
	\end{align}
	due to $\log\Big(\lambda  \phi(Z)\Big)\leq \log\frac{\lambda}{\sqrt{2\pi\sigma^2}}<0$ which holds for typical values of $\sigma$ we need to considered.
	We show that the second term of \eqref{eq:Efai3} is less than $\Big(\frac{3\Delta R^4}{4\sigma^4}+\frac{3}{2\Delta^2}\Big)e^{-\frac{(\Delta R-\beta)^2}{8\sigma^2}}$ where $\beta=\frac{2\sigma^2}{\Delta R}\log\frac{1-\Delta}{\Delta}$. 
	
	Let $\tau=\frac{\Delta R}{2}+\frac{\sigma^2}{\Delta R}\log\frac{\lambda}{1-\lambda}$. 
	Using the similar technique as in Appendix~\ref{app:approx},
	we have
	\begin{align}
	&E\Bigg[\Bigg(\log\Bigg(1+\frac{1-\lambda }{\lambda }e^{-\frac{\Delta R(\Delta R-2Z)}{2\sigma^2}}\Bigg)\Bigg)^2\Bigg]\nonumber\\ \leq &\ E\Bigg[\frac{(1-\lambda)^2}{\lambda^2 }e^{-\frac{\Delta R(\Delta R-2Z)}{\sigma^2}}1\Big\{Z\leq\tau\Big\}\Bigg]\nonumber \\
	&   \  +E\Bigg[\Big(1+\log \frac{1-\lambda }{\lambda }-\frac{\Delta R(\Delta R-2Z)}{2\sigma^2}\Big)^21\Big\{Z>\tau\Big\}\Bigg]\label{eq:Evv2}.
	\end{align}
	The first term of the above expression is bounded by 
	\begin{align}
	E\Bigg[\frac{(1-\lambda)^2}{\lambda^2 }e^{-\frac{\Delta R(\Delta R-2Z)}{\sigma^2}}1\Big\{Z\leq\tau\Big\}\Bigg]&=\frac{(1-\lambda)^2}{\lambda^2}e^{\frac{\Delta R^2}{\sigma^2}}Q\Big(\frac{2\Delta R-\tau}{\sigma}\Big)
	\nonumber\\
	&\leq \frac{(1-\lambda)^2}{2\lambda^2}e^{-\frac{(2\Delta R-\tau)^2-2\Delta R^2}{2\sigma^2}}\nonumber\\
	&\leq  \frac{1}{2\Delta^2}e^{-\frac{\tau^2}{2\sigma^2}}\label{eq:firstE2}
	\end{align}
	where we have used $\frac{(1-\lambda)^2}{2\lambda^2}\leq \frac{1}{2\Delta^2}$ and $\tau\leq \Delta R$.	
	
	To evaluate the second term of \eqref{eq:Evv2}, we first have
	\begin{align}
	&\ \Bigg(1+\log\frac{1-\lambda}{\lambda }-\frac{\Delta R^2-\Delta R Z}{2\sigma^2}\Bigg)^2\nonumber\\
	\leq&\ 2\Bigg(1+\log\frac{1-\lambda}{\lambda }\Bigg)^2+2\Bigg(\frac{\Delta R^2-\Delta R Z}{2\sigma^2}\Bigg)^2\nonumber\\
	\leq& \frac{2}{\Delta^2} + \frac{\Delta R^4+\Delta R^2Z^2}{2\sigma^4}\label{eq:a2b2}
	\end{align}
	where we have used $\big|1+\log\frac{1-\lambda}{\lambda }\big|\leq \frac{1}{\Delta}$ and the condition $Z>\tau>0$.
	Moreover, since $\tau<\Delta R$ and $\sigma<\Delta R$, we have
	\begin{equation}
	E\Big[1\{Z>\tau\}Z^2\Big]=\frac{\sigma\tau}{\sqrt{2\pi}}e^{-\frac{\tau^2}{2\sigma^2}}+\sigma^2Q\big(\frac{\tau}{\sigma}\big)  \leq \Delta R^2e^{-\frac{\tau^2}{2\sigma^2}}\label{eq:Q2}.
	\end{equation}
	Utilizing \eqref{eq:Q0}, \eqref{eq:a2b2}, and \eqref{eq:Q2}, we obtain that the second term of \eqref{eq:Evv2} is upper bounded by  
	\begin{align}
	&E\Bigg[\Big(1+\log \frac{1-\lambda }{\lambda }-\frac{\Delta R(\Delta R-2Z)}{2\sigma^2}\Big)^21\Big\{Z>\tau\Big\}\Bigg]\nonumber\\
	\leq &\ \Big(\frac{1}{\Delta^2}+\frac{3\Delta R^4}{4\sigma^4}\Big)e^{-\frac{\tau^2}{2\sigma^2}}\label{eq:secondE2}.
	\end{align}
	Applying \eqref{eq:firstE2} and \eqref{eq:secondE2} to \eqref{eq:Evv2}, we obtain 
	\begin{align}
	E\Bigg[\Bigg(\log\Bigg(1+\frac{1-\lambda }{\lambda }e^{-\frac{\Delta R(\Delta R-2Z)}{2\sigma^2}}\Bigg)\Bigg)^2\Bigg]&\leq \Big(\frac{3\Delta R^4}{4\sigma^4}+\frac{3}{2\Delta^2}\Big)e^{-\frac{(\Delta R-\beta)^2}{8\sigma^2}}\nonumber
	\end{align}
	where we have used $\tau\geq(\Delta R-\beta)/2$.

\end{document}